\documentclass[pra,superscriptaddress,showpacs,twocolumn]{revtex4}

\usepackage{amsmath,amssymb,graphicx,color}

\newcommand{\ket}[1]{|#1\rangle}
\newcommand{\bra}[1]{\langle#1|}
\newcommand{\olap}[2]{\langle#1|#2\rangle}
\newcommand{\B}{\hat{\mathcal{B}}}

\newcommand{\be}{\begin{equation}}
\newcommand{\ee}{\end{equation}}
\newcommand{\ba}{\begin{eqnarray}}
\newcommand{\ea}{\end{eqnarray}}
\newcommand{\ban}{\begin{eqnarray*}}
\newcommand{\ean}{\end{eqnarray*}}

\newcommand{\braket}[2]{\mbox{$ \langle #1 | #2 \rangle $}}

\newcommand{\Tr}{\mbox{Tr}}

\newcommand{\proj}[2]{\ket{#1}\bra{#2}}

\newcommand{\demi}{\frac{1}{2}}
\newcommand{\compl}{\begin{picture}(8,8)\put(0,0){C}\put(3,0.3){\line(0,1){7}}\end{picture}}

\newcommand{\one}{\leavevmode\hbox{\small1\normalsize\kern-.33em1}}

\begin{document}

\title{Device independent state estimation based on Bell's inequalities}
\author{C.-E. Bardyn}
\affiliation{Centre for Quantum Technologies, National University of
Singapore, Singapore 117543} \affiliation{Ecole Polytechnique
F\'{e}d\'{e}rale de Lausanne (EPFL), 1015 Lausanne, Switzerland}
\author{T. C. H. Liew}
\affiliation{Centre for Quantum Technologies, National University of
Singapore, Singapore 117543}
\author{S. Massar}
\affiliation{Laboratoire d'Information Quantique, CP2 225, Universit\'e Libre
de Bruxelles, Av. F. D. Roosevelt 50, B-1050 Belgium}
\author{M. McKague}
\affiliation{Institute for Quantum Computing and Department of
Combinatorics \& Optimization, University of Waterloo, N2L 3G1
Canada}
\author{V. Scarani}
\affiliation{Centre for Quantum Technologies, National University of
Singapore, Singapore 117543} \affiliation{Department of Physics,
National University of Singapore, Singapore 117542}
\date{\today}

\begin{abstract}
The only information available about an alleged source of entangled quantum states is the amount $S$ by which the Clauser-Horne-Shimony-Holt (CHSH) inequality is violated: nothing is known about the nature of the system or the measurements that are performed. We discuss how the quality of the source can be assessed in this black-box scenario, as compared to an ideal source that would produce maximally entangled states (more precisely, any state for which $S=2\sqrt{2}$). To this end, we introduce several inequivalent notions of fidelity, each one related to the use one can make of the source after having assessed it; and we derive quantitative bounds for each of them in terms of the violation $S$. We also derive a lower bound on the entanglement of the source as a function of $S$ only.
\end{abstract}


\maketitle

\section{Introduction}

\begin{figure}[ht]
\includegraphics[scale=0.5]{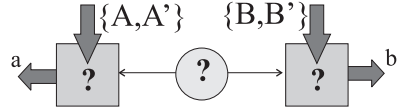}
\caption{Device-independent state estimation: the quality of an unknown source of entangled pairs should be established using unknown measurement devices. The only available information are the statistics $P(a,b|A,B)$ of the outcomes $(a,b)$ for measurement settings $A,B$. The figure represents the particular case studied in this paper, where both the choice of measurement settings and the outputs are binary.}
\label{fig:scheme}
\end{figure}

A device, allegedly generating pairs of entangled particles, is for sale. Obviously, the potential \textit{user} wants to check that entanglement is indeed being generated before buying it; but just as obviously, the \textit{vendor} does not want to open the device and reveal its fabrication. For classical devices, such a situation would lead to a complete impasse. Not so, however, for quantum devices: \textit{Bell's inequalities} can act as entanglement witnesses irrespective of the nature of the system under study or of the kind of measurements that are being performed. Thus suppose that the vendor provides the user with two additional boxes, the measurement devices. Once more the vendor does not want to open the device and reveal its fabrication. Suppose in addition that the user can choose the measurements: the measurement devices have a knob whose positions correspond to allegedly different measurements (Fig.~\ref{fig:scheme}). By operating these devices, the user can reconstruct the statistics $P(a,b|A,B)$ of the observed outputs $a$ and $b$, conditioned on each choice of knob positions $A$ and $B$. If the statistics violate some Bell inequality, and the measurement has been performed in such a way as to avoid signaling between the measurement boxes, then the user is convinced that the source is indeed producing entangled pair.

The possibility of such an assessment is already remarkable. However, the user cannot be satisfied with knowing that there is ``some entanglement'': what is needed is a \textit{quantitative estimate} on how good the source actually is. The amount of violation of a Bell's inequality can provide such a quantitative criterion, provided it is translated into the meaningful figure of merit: fidelity or trace distance to the ideal state, or some entanglement measure... The goal of this paper is to provide such quantitative estimates, when the Bell inequality under study is the CHSH inequality derived by Clauser, Horne, Shimony and Holt \cite{Clauser1969}.

This work is inspired by ``device-independent quantum key distribution'' \cite{Acin2007,lemma}, in which the amount of violation of the CHSH inequality is used to bound the information of an eavesdropper without making any hypothesis on the internal workings of the devices. It is also related to the concept of ``dimension witness'': sufficient violation of some Bell inequalities can guarantee that the quantum state has a minimum dimension \cite{Brunner2008,Vertesi2008,Briet2009}. One possible application of the present work could be to devise improved 
self testing of quantum computers \cite{Mayers2004,Magniez06}.

\section{Formulation of the problem}

\subsection{
Ideal states}

As we said, we restrict to the case where the user applies only two measurement settings on each particle and the outcome is binary. In this case, there is only one Bell inequality, namely CHSH \cite{fine}. We further restrict our study in considering only the observed violation $S_{\mathrm{obs}}$ of CHSH as quantitative measure, being aware that the statistics $P(a,b|A,B)$ contain further information that might improve the estimates.

Since the source will be characterized by a single scalar quantity, the set of ideal states is the set of states $\boldsymbol{\Phi}$ such that $S=2\sqrt{2}$ is achievable. This set has been fully characterized \cite{bmr92,pr92}: 
it consists of all pure states of the form $\sum_{j}c_j\ket{\Psi_j}$ where $\ket{\Psi_j}$ is a two-qubit maximally entangled state in a four-dimensional subspace, i.e., $\ket{\Psi_j}=\frac{1}{\sqrt{2}}(|2j-1,2j-1\rangle+|2j,2j\rangle)$ up to local unitaries. Since the relative phases of the $c_j$ do not play any role in the violation, we must add mixed states to the set. It is easy to verify that the most general such state can be written as 
$\boldsymbol{\Phi}=U_A U_B \boldsymbol{\Phi^+} \otimes \boldsymbol{\sigma} U_A^\dagger U_B^\dagger$ where $\ket{\Phi^+}=\frac{1}{\sqrt{2}}(|00\rangle+|11\rangle)$ is a two-qubit maximally entangled state, $\boldsymbol{\sigma}$ is an arbitrary state, and $U_A$, $U_B$ are arbitrary local unitaries.
In their work on device-testing, Mayers and Yao (MY) \cite{Mayers2004} chose their reference states as those 
that could be written in the above form with $\boldsymbol{\sigma}$ pure, i.e., even though they did not refer to Bell inequalities, they where considering all pure states that violate CHSH maximally.

\subsection{Figures of merit}

The distance between the actual source state, with density matrix $\boldsymbol{\rho}$, and the closest ideal state $\boldsymbol{\Phi}$, is conveniently measured by the trace distance~\cite{Fuchs1999,Nielsen2000}:
\be\label{deltaMY}
\delta_{MY}(\boldsymbol{\rho})=
\min_{\boldsymbol{\Phi}} \delta(\boldsymbol{\rho},\boldsymbol{\Phi})\,,
\ee
where $\delta(\boldsymbol{\rho},\boldsymbol{\Phi})=\frac{1}{2}\,\Tr |\boldsymbol{\rho}-\boldsymbol{\Phi}|$. The trace distance has a clear operational interpretation: in whatever task, $\boldsymbol{\rho}$ will behave differently from $\boldsymbol{\Phi}$ with probability at most $\delta(\boldsymbol{\rho},\boldsymbol{\Phi})$. In other words, the real source will differ from an ideal source with probability at most $\delta(\boldsymbol{\rho})$.

The problem we have set out to solve is thus to find a bound of the form
\be\label{DMY}
\delta_{MY}(\boldsymbol{\rho})\leq {\cal D}_{MY}(S_{\mathrm{obs}})\,.
\ee
This bound can in principle be obtained by solving the following optimization problem:
\begin{equation}
{\cal D}_{MY}(S_{\mathrm{obs}})=\max_{\boldsymbol{\rho}:S_{\mathrm{max}}(\boldsymbol{\rho})\geq S_{\mathrm{obs}}}\left\{\min_{\boldsymbol{\Phi}}\delta(\boldsymbol{\rho},\boldsymbol{\Phi})\right\}\,,\label{eq:D}
\end{equation} 
where $S_{\mathrm{max}}(\boldsymbol{\rho})$ is the maximum CHSH violation that can be obtained by measuring state $\boldsymbol{\rho}$. 

Deriving lower bounds, let alone tight lower bounds, for ${\cal D}_{MY}$ turns out to be much harder than we initially anticipated. In practice, it is simpler to work with the \textit{Fidelity} rather than the trace distance, the two measures being related by $\delta \leq \sqrt{1-F}$ \cite{Nielsen2000}. In analogy with eq.~(\ref{deltaMY}), we define 
\be\label{FMY}
F_{MY}(\boldsymbol{\rho})= \max_{\boldsymbol{\Phi}} F(\boldsymbol{\rho},\boldsymbol{\Phi})=
\max_{\boldsymbol{\Phi}} \left(\Tr\sqrt{\boldsymbol{\rho}^{1/2}\boldsymbol{\Phi}\boldsymbol{\rho}^{1/2}}\right)^2\,
\ee
and one is then led to search for bounds of the form
\be\label{FFMY}
F_{MY}(\boldsymbol{\rho})\geq {\cal F}_{MY}(S_{\mathrm{obs}})\,.
\ee
For ${\cal F}_{MY}$ we obtain tight lower bounds if the state is restricted to consist of two qubits, or (modulo a conjecture of Gisin and Peres) if the state is restricted to be pure. Putting such hypotheses on the source goes against the philosophy of the black box scenario, but it allows us to get a mathematical grasp of the problem. When no restrictions are put on the state, we do not even have a lower bound on ${\cal F}_{MY}$ . However it is possible to introduce other notions of fidelity (see below) which have a clear operational meaning, and for which lower bounds can be computed without any hypothesis on the source.

Yet an alternative approach to the source characterization problem would consist in looking for a lower bound to the \textit{entanglement} of the state $\boldsymbol{\rho}$:
\be\label{Ent}
E(\boldsymbol{\rho})\geq {\cal E}(S_{\mathrm{obs}})\,,
\ee
where $E$ is an entanglement measure, such as the entanglement of formation, of distillation, etc... \cite{H4}. Below we obtain lower bounds on ${\cal E}$.

\subsection{Warm-up: solution assuming two qubits}

As a nontrivial warm up exercise, let us compute the bound eq. (\ref{FFMY}) under the assumption that the source emits a pair of qubits and that the measurements are von Neumann measurements. This is an undue restriction for the black-box scenario; we present this calculation because its result is interesting in itself, and will be an important tool for the main discussion. 

In this case the set of ideal states is well known: only the maximally entangled states
$\boldsymbol{\Phi}= U_AU_B \boldsymbol{\Phi^+}U_A^\dagger U_B^\dagger$ violate CHSH maximally. Therefore, $F_{MY}(\boldsymbol{\rho})=\max_{\boldsymbol{\Phi}} F(\boldsymbol{\rho},\boldsymbol{\Phi})$ reduces to the so called \textit{singlet fidelity} of $\boldsymbol{\rho}$. Our approach consists in fixing the singlet fidelity of $\boldsymbol{\rho}$, and computing $S_{\mathrm{max}}(\boldsymbol{\rho})$. To this end we use the spectral decomposition of the Bell operator
\begin{align}
\B&=\big(\hat{A}+\hat{A}^\prime\big)\otimes \hat{B}+\big(\hat{A}-\hat{A}^\prime\big)\otimes \hat{B}^\prime\label{eq:Bell}\,.
\end{align}
First note that if $F(\boldsymbol{\rho})\leq\demi$ the state cannot be entangled, CHSH cannot be violated, and the bound $S_{\mathrm{max}}=2$ can be trivially achieved by the degenerate measurement $\hat{A}=\hat{A}^\prime=\hat{B}=\hat{B}^\prime=\one$. If the inequality is violated, the operators $\hat{A}$, $\hat{A}^\prime$, $\hat{B}$, and $\hat{B}^\prime$ must be linear combinations of the three Pauli matrices. Then the spectral decomposition $\B=\sum_i \lambda_i \ket{\Phi_i}\bra{\Phi_i}$ has the following properties \cite{Scarani2001}: the $\ket{\Phi_i}$ are a Bell basis (i.e., a basis of maximally entangled states) and the eigenvalues are $\{\lambda_1,\lambda_2,-\lambda_2,-\lambda_1\}$ with $\Tr(\B^2)=16$, i.e., 
\begin{equation}
\lambda_1^2+\lambda_2^2=8\,,
\label{eq:2Qubitconstraint}
\end{equation}
which implies the Cirelson bound $|\lambda_i|\leq2\sqrt{2}$ \cite{Cirelson1980}.

Therefore, for a given $\B$ we have
\begin{equation}
S(\boldsymbol{\rho})=\Tr(\boldsymbol{\rho}\B)=\sum_i \lambda_i \bra{\Phi_i}\boldsymbol{\rho}\ket{\Phi_i}\,.
\end{equation} 
Suppose for definiteness $\lambda_1\geq\lambda_2\geq 0$. Then, keeping $F(\boldsymbol{\rho})$ fixed,
$S(\boldsymbol{\rho})$ is maximized by choosing $\ket{\Phi_1}$ such that $F(\boldsymbol{\rho},\boldsymbol{\Phi_1})=F(\boldsymbol{\rho})$. Whereupon we have that $S_{\mathrm{max}}(\boldsymbol{\rho})\leq \lambda_1 F(\boldsymbol{\rho})+\lambda_2(1-F(\boldsymbol{\rho}))$ because the two other eigenvalues are non-positive. Using eq.~(\ref{eq:2Qubitconstraint}), we can set $\lambda_1=2\sqrt{2}\cos x$ and $\lambda_2=2\sqrt{2}\sin x$. The well-known bound $\max_x(a\cos x+b\sin x)=\sqrt{a^2+b^2}$ then leads to
$S_{\mathrm{max}}(\boldsymbol{\rho})\leq 2\sqrt{2}\sqrt{F(\boldsymbol{\rho})^2+[1-F(\boldsymbol{\rho})]^2}$. Finally (Fig.~\ref{fig:curves}):
\ba
\label{eq:FS}
&F_{MY}(\boldsymbol{\rho})\geq \left(1+\sqrt{[S_{\mathrm{obs}}/2]^2-1}\right)/2
\quad \mbox{(qubits).}&
\ea 
This bound is tight, being achieved by pure non-maximally entangled states $\ket{\psi}=\cos\theta\ket{00}+\sin\theta\ket{11}$. Indeed, for these states $S_{\mathrm{max}}=2\sqrt{1+\sin^2(2\theta)}$ \cite{Werner2001,Horodecki1995} and the singlet fidelity is $F=\left|\olap{\psi}{\Phi^+}\right|^2=\demi(1+\sin(2\theta))$. Furthermore for pairs of pure states, we have the strict equality $\delta=\sqrt{1-F}$, hence eq.~(\ref{eq:FS}) leads to a tight bound for the trace distance as well.

\section{Bounds on the fidelity to the closest reference state}

\subsection{Structure of the Bell operator}

For any two dichotomic operators $\hat{A}$ and $\hat{A}^\prime$, one can find a basis such that both operators are block-diagonal, where each block is a $2\times 2$ matrix (see e.g. \cite{lemma}). So one has $\hat{A}=\sum_{\alpha}\hat{A}_\alpha$ and $\hat{A}^\prime=\sum_{\alpha}\hat{A}^\prime_\alpha$ where $\hat{A}_\alpha=\Pi_\alpha \hat A \Pi_\alpha$, 
$\hat{A}^\prime_\alpha=\Pi_\alpha \hat A^\prime \Pi_\alpha$
and $\Pi_\alpha$ are orthogonal projectors onto 2 dimensional spaces.
Of course, a similar decomposition holds for Bob's operators. Therefore, the Bell-CHSH operator can be written as
\be\label{Bab}
\B=\sum_{\alpha,\beta} \B_{\alpha,\beta}\,,
\ee
where $\B_{\alpha,\beta}=\sum_i \lambda_{i}^{\alpha\beta}\ket{\Phi_{i}^{\alpha\beta}} \bra{\Phi_{i}^{\alpha\beta}}$ are orthogonal two-qubit operators with the same properties as above. Therefore
\begin{equation}
S(\boldsymbol{\rho})=\sum_{\alpha,\beta}p_{\alpha\beta}\Tr(\boldsymbol{\rho}_{\alpha\beta}\B_{\alpha\beta})\,=\, \sum_{\alpha,\beta}p_{\alpha\beta}S(\boldsymbol{\rho}_{\alpha\beta})\,,\label{Sdim}
\end{equation}
where $p_{\alpha\beta}\boldsymbol{\rho}_{\alpha\beta}=
\Pi_\alpha \otimes \Pi_\beta {\boldsymbol{\rho}} \Pi_\alpha \otimes \Pi_\beta$ and $\boldsymbol{\rho}_{\alpha\beta}$ is a normalized two-qubit state.

\subsection{A complex problem}

Given (\ref{Sdim}), it may seem that the extension of our result to arbitrary dimensions is just a matter of convex optimization. A closer look shows that one must be much more careful, because the above construction does not imply:
\ba
F_{MY}(\boldsymbol{\rho})&\geq & \sum_{\alpha,\beta}p_{\alpha\beta}F(\boldsymbol{\rho}_{\alpha\beta})\quad\mbox{(probably wrong),}
\label{Fwrong}
\ea
where $F(\boldsymbol{\rho}_{\alpha\beta})$ is the singlet fidelity of $\boldsymbol{\rho}_{\alpha\beta}$. The reason is that in the MY approach, the state must be brought close to a reference state using local unitary operations $U_A\otimes U_B$. Let $U_{\alpha}$ be the restriction of $U_A$ to the $2\times 2$ block indexed by $\alpha$; and similarly for $U_{\beta}$; and let $\boldsymbol{\Phi}_{\alpha\beta}$ be the maximally entangled state of two qubit such that $F(\boldsymbol{\Phi}_{\alpha\beta},\boldsymbol{\rho}_{\alpha\beta})=F(\boldsymbol{\rho}_{\alpha\beta})$ is the singlet fidelity of $\boldsymbol{\rho}_{\alpha\beta}$. Now, there is no guarantee that $U_A$ and $U_B$ exist, such that $U_\alpha\otimes U_\beta\boldsymbol{\Phi}_{\alpha\beta}U_\alpha^\dagger\otimes U_\beta^\dagger=\boldsymbol{\Phi}^+$ for all $\alpha$ and $\beta$, as is required to obtain a reference state according to the MY definition. Moreover, the MY definition of fidelity is a comparison with the whole state $\boldsymbol{\Phi}^+\otimes\boldsymbol{\sigma}$, not only with the two-qubit component $\boldsymbol{\Phi}^+$. In order to make sense of eq. (\ref{Fwrong}) we will introduce different definitions of fidelity below. Before turning to that, we present the case of pure states of arbitrary dimensions, for which the MY fidelity can be computed.

\subsection{Solution under the restriction to pure states}

\begin{figure}[ht]
\includegraphics[scale=0.40]{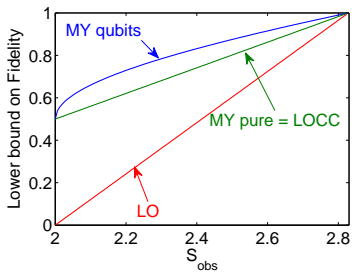}
\caption{(Color online) Lower bounds on the fidelity as a function of the observed violation $S_{\mathrm{obs}}$ of the CHSH inequality. From top to bottom: $F_{MY}$ assuming two qubits (\ref{eq:FS}); $F_{MY}$ assuming pure states and the Gisin-Peres conjecture(\ref{fidbound}), equal to  $F_{LOCC}$ (\ref{fidboundLOCC}); and $F_{LO}$ (\ref{fLObound}).}
\label{fig:curves}
\end{figure}

Let us assume that we know that the source emits a pure state (again an undue restriction for the black box scenario). Using the Schmidt decomposition $\ket{\Psi}=\sum_k \lambda_k\ket{k,k}$ with the Schmidt coefficients in decreasing order $\lambda_k\geq\lambda_{k+1}\geq 0$, any pure state can be rewritten as $\ket{\Psi}= \sum_{j}\sqrt{p_j}\left(c_j\ket{2j,2j}+s_j\ket{2j+1,2j+1}\right)$ with $c_j^2+s_j^2=1$. The MY fidelity can be computed exactly (Appendix \ref{appa}), and one finds
\be\label{eq:FMYpure}
F_{MY}(\Psi)=\sum_j \frac{(\lambda_j + \lambda_{j+1})^2}{2}=\sum_j p_j \frac{(c_j+s_j)^2}{2}\,.\ee 
This should now be related to 
$S_{\mathrm{max}}(\Psi)$. For states of arbitrary dimension, there is no known analytical expression for the maximal violation of CHSH. However, for pure states $\Psi$ there is a long-standing conjecture by Gisin and Peres \cite{Gisin1992}, whose validity has never been disproved by numerical checks \cite{Liang2006}. According to this conjecture, the ordered Schmidt decomposition defines the natural block-structure of the CHSH operator. This implies 
\be
S_{\mathrm{max}}(\Psi)=\sum_jp_j\left[2\sqrt{1+4c_j^2s_j^2}\right]\,.
\label{SGisinPeres}
\ee 
Combining this conjecture with eq.~(\ref{eq:FMYpure}) we find that for pure states the accessible points in the $( F_{MY}, S_{\mathrm{max}})$ plane are convex combinations of points on the curve given by equality in eq.~(\ref{eq:FS}), yielding (Fig.~\ref{fig:curves}):
\ba 
&F_{MY}(\boldsymbol{\Psi})\geq \frac{1}{4(\sqrt{2}-1)}\left[S_{\mathrm{obs}}+2\sqrt{2}-4\right]& \label{fidbound}
\\
&\mbox{(pure states, modulo Gisin-Peres conjecture).}\nonumber
\ea
This bound is tight if we allow the dimension $d$ to become arbitrarily large (otherwise, the ordering of the $\lambda_k$ implies constraints on the possible values of $\{p_j,c_j,s_j\}$). Moreover, this bound is weaker than the one obtained under the assumption of two-qubits. Though not astonishing in itself, this feature is new: in device-independent quantum key distribution, the bound for collective attacks is already optimal in the two-qubit case \cite{Acin2007,lemma}.

\subsection{Black-box bounds for other fidelities}

The MY fidelity is defined to suit the black-box scenario. However, other definitions of fidelity may be meaningful. Here, we consider fidelities defined as \be
 	F_{{\cal L}}(\boldsymbol{\rho}) = \max_{\Lambda \in {\cal L}} F(\Lambda(\boldsymbol{\rho}), \boldsymbol{\Phi^+})\,,\label{eq:Fnew}
\ee
where ${\cal L}$ is a set of completely positive maps which map the Hilbert space of $\boldsymbol{\rho}$ onto a $2\times 2$ dimensional Hilbert space, and which cannot increase the entanglement. $F_{\cal L}$ can be thought of as the best singlet fidelity obtainable under single shot purification of $\boldsymbol{\rho}$ to a two-qubit entangled state, using only operations that belong to the family ${\cal L}$.
We will consider the case where ${\cal L}$ consists of all the completely positive maps that can be realised by Local Operations (${\cal L}=LO$) or by Local Operations and Classical Communication (${\cal L}=LOCC$). 

These new notions of fidelity shed a different light on the task of source characterization. Indeed the Mayers-Yao fidelity and trace distance compare the state produced by the source to the closest ideal state, thereby establishing how much the real state and the idea state would differ in applications. The new fidlities $F_{\cal L}$ are relevant to another scenario in which the user may try to improve the source by acting locally on the two subsystems, for instance by opening the boxes containing the measurement devices and tinkering inside them. But before buying the source, the user wants to perform a fast black-box check to ascertain what will be the performance of the improved source. In other words, by measuring $S_{\mathrm{obs}}$ in a black-box scenario, the user can assess how well the source would perform in other scenarios. In this sense, the bounds we derive for these new fidelities are real black box statement, which do not make any hypothesis on the state $\boldsymbol{\rho}$ and on the measurement devices.

These new fidelities are related by
\ba
F_{MY}\leq F_{LO}\leq F_{LOCC}\,.
\label{inequ}
\ea
In Appendix \ref{appb}, we prove that $F_{MY}(\boldsymbol{\Psi})=F_{LO}(\boldsymbol{\Psi})$ for pure states while for mixed states there are explicit cases of strict inequality. 
We also show, see Fig.~\ref{fig:curves}, that:
\ba 
F_{LO}(\boldsymbol{\rho})&\geq& \frac{1}{2(\sqrt{2}-1)}\left[S_{\mathrm{obs}}-2\right]\,,\label{fLObound}\\
F_{LOCC}(\boldsymbol{\rho}) &\geq& \frac{1}{4(\sqrt{2}-1)}\left[S_{\mathrm{obs}}+2\sqrt{2}-4\right]\,.\label{fidboundLOCC}
\ea
The bound (\ref{fLObound}) on $F_{LO}$ is obtained by exhibiting an explicit LO strategy. The proof is lengthy and we give it in Appendix \ref{appc}. We just note here that this bound is surely not tight, since it reaches the over-pessimistic $F=0$ for $S=2$.

The bound (\ref{fidboundLOCC}) for $F_{LOCC}$ is the same one obtained for $F_{MY}$ on pure states, eq.~(\ref{fidbound}); we do not know whether this bound is tight. The proof goes as follows. The decomposition eq.~(\ref{Bab}) of the Bell-CHSH operator gives us a natural method for projecting $\boldsymbol{\rho}$ onto a 2 qubit space: particle A is projected in the $\Pi_\alpha$ spaces and particle B in the $\Pi_\beta$ spaces. Using CC, the actual block $(\alpha,\beta)$ is made known in both locations. The result of this completely positive map is a state with fidelity $F=p_{\alpha \beta}\sum_{\alpha \beta}F(\boldsymbol{\rho}_{\alpha \beta})$, i.e., we obtain a convex combination of points on the curve eq.~(\ref{eq:FS}). The bound (\ref{fidboundLOCC}) is recovered by noticing that, in the LOCC scenario, all the blocks $(\alpha,\beta)$ for which $S(\boldsymbol{\rho}_{\alpha \beta})=2$ can be brought to have $F=1/2$: indeed, for the blocks where they observe $S=2$, Alice and Bob can swap their local states with those of ancillas prepared in a pure product state.

\section{Other figures of merit}

The core of our work involved using the fidelity as a figure of merit. Here, we present the consequences of the bounds obtained on the fidelity for other figures of merit.

\subsection{Relation with trace distance}

Even if fidelity bounds were found to be tight, the tightness of the bound $\delta\leq\sqrt{1-F}$ on the trace distance would follow only if the states that saturate the bound are pure. However, we are already able to conclude that the bounds ${\cal D}(S_{\mathrm{obs}})$ for the trace distance $\delta$ put very stringent constraints on the quality of the source.

For instance, our strongest bound eq.~(\ref{eq:FS}) leads to a tight $\delta=\sqrt{1-F}$. If we insert $S_{\mathrm{obs}}=0.99\times 2 \sqrt{2}$, we obtain $\delta\approx 10\%$. If the user requests the error rate to be below 1\%, the vendor will have to produce extremely good sources --- better than any currently available one.

\subsection{Relation with entanglement measures}

The bounds for all the ${\cal F}_{\cal L}$ also provide lower bounds on the entanglement of $\boldsymbol{\rho}$. Indeed, consider any entanglement measure $E$ (see \cite{H4} for a list). By definition, ${\cal L}$ is a set of operations under which $E$ cannot increase; and the bounds on ${\cal F}_{\cal L}$ tell us how close the state $\boldsymbol{\rho}$ can be brought to the singlet state using only operations in ${\cal L}$. If ${\cal L}=LOCC$, each $\boldsymbol{\rho}_{\alpha\beta}$ can further be twirled, leading to the map $\boldsymbol{\boldsymbol{\rho}} \to p\, \boldsymbol{\Phi^+} + (1-p) \frac{\one}{4}$ with $p=\left( 4 {\cal F}_{LOCC}(S_{\mathrm{obs}}) +1\right)/3$.
For such states the entanglement measures can generally be computed. For instance, using \cite{Bennet96}, the entanglement of formation is bounded by
$$
E_f\geq h\left(\frac{1}{2} +\frac{1}{4(\sqrt{2}-1)}\sqrt{8(1-\sqrt{2}) + 4 S_{\mathrm{obs}} - S^{2}_{\mathrm{obs}}} \right).$$
where $h$ is the binary entropy function.

\section{Conclusion}

A theory of black box source characterization is a step towards the development of device-independent quantum information processing. In the present work we used only the CHSH inequality: already in this simple case, we have uncovered a rich structure, raised many problems and solved a few.

In particular, the task of deriving black-box bounds for use in the black-box scenario in full generality is still open; we have been able to derive tight bounds for the Mayers-Yao fidelity either by restricting the dimensions to two qubits (\ref{eq:FS}), or by restricting the state to be pure (\ref{fidbound}). For arbitrary states, we do not even have a lower bound for the Mayers-Yao fidelity or trace distance. However we have been able to derive unrestricted black-box bounds for use in other scenarios (\ref{fLObound},\ref{fidboundLOCC}) where one wants to ascertain how close to an ideal state it would be possible to bring the system by local operations, possibly complemented by classical communication. We have also been able to derive unrestricted black-box lower bounds for the entanglement of the state. 

Our results indicate that black-box bounds put very stringent demands on the quality of an untrusted source, which could in particular have important consequences for self testing of quantum computers.

\section*{Acknowledgments}

This work is supported by the National Research Foundation and Ministry of Education, Singapore; by the Interuniversity Attraction Poles (Belgian Science Policy) project IAP6-10 Photonics@be; by the EU projectQAP contract 015848; by NSERC, Ontario-MRI, OCE, QuantumWorks, MITACS, and the Government of Canada. We are grateful to Fr\'ed\'eric Magniez, Mike Mosca and Jamie Sikora for helpful discussions.

\appendix

\section{Calculating $F_{MY}$ for pure states}\label{appa}

We begin with a state $\ket{\psi}$ in Schmidt form
\begin{equation}
\ket{\psi} = \sum_{j} \lambda_{j} \ket{a_{j}}\ket{b_{j}}
\end{equation}
while the closest state of the form $\ket{?} \otimes \ket{\phi_{+}}$ has Schmidt decomposition
\begin{equation}
\ket{\phi} = \sum_{j} \mu_{j} \ket{c_{j}}\ket{d_{j}}.
\end{equation}
with $\mu_{2l} = \mu_{2l+1}$.  For concreteness, we may assume that the $\lambda_{j}$s and $\mu_{j}$s are both in decreasing order

We first show that we may take $\ket{c_{j}} = \ket{a_{j}}$ and $\ket{d_{j}} = \ket{b_{j}}$.  Note that
\begin{equation}
|\braket{\psi}{\phi}| \leq \sum_{jk} \lambda_{j} \mu_{k} |\braket{a_{j}}{c_{k}}||\braket{b_{j}}{d_{k}}|
\end{equation}
Let us define the matrix $M$ by 
\begin{equation}
M_{jk} = |\braket{a_{j}}{c_{k}}||\braket{b_{j}}{d_{k}}|.
\end{equation}
The values $|\braket{a_{j}}{b_{k}}|$ for various $k$ and fixed $j$ form a vector of norm 1 since $\ket{b_{k}}$ is a basis and $\ket{a_{j}}$ has norm 1.  The same is true for the values $|\braket{b_{j}}{d_{k}}|$ and if we fix $k$ and vary $j$ instead.  Thus columns (and rows) of $M$ are formed by entrywise products of norm 1 vectors and the sum of each row and column of $M$ is at most 1.  This means that we can find a new matrix $N$ with positive entries such that $M + N$ is doubly stochastic.  Note that
\begin{equation}
|\braket{\psi}{\phi}| \leq \sum_{jk} \lambda_{j} \mu_{k} (M+N)_{jk}.
\end{equation}

By the Birkhoff-von Neumann theorem we may write $M+N$ as a convex combination of permutation matrices, thus
\begin{equation}
M+N = \sum_{m} p_{m} P_{m}
\end{equation}
with $\sum_{m} p_{m} = 1$ and $P_{m}$ permutation matrices.  Since the combination is convex, there exists some $m$ for which
\begin{equation}
|\braket{\psi}{\phi}| \leq \sum_{jk}\lambda_{j} \mu_{k} (P_{m})_{jk}.
\end{equation}
The permutations merely reorder the $\mu_{j}$s and it is easy to prove that the maximum is achieved when the $\lambda_{j}$s and $\mu_{j}$s are both in decreasing order.  Hence $P_{m} = I$ satisfies the above equation.  We may achieve this by choosing the bases $\ket{c_{j}} = \ket{a_{j}}$ and $\ket{d_{j}} = \ket{b_{j}}$, so we need not consider any other bases. 

We now optimize over $\mu_{j}$ subject to the condition $\mu_{2l} = \mu_{2l+1}$.  By the Cauchy-Schwarz inequality we have

\begin{equation*}
|\braket{\psi}{\phi}|^{2} = \left(\sum_{l} (\lambda_{2l} + \lambda_{2l + 1})\mu_{2l}\right)^{2}
\end{equation*}
\begin{equation} \leq 
\left(\sum_{l} (\lambda_{2l} + \lambda_{2l + 1})^{2}\right) \left(\sum_{l} \mu_{2l}^{2} \right)
\end{equation}

with equality when $\mu$ and $\lambda$ are collinear.  Thus we set
\begin{equation}
\mu_{2l} = \mu_{2l+1} = \frac{\lambda_{2l} + \lambda_{2l+1}}{N}
\end{equation}
with $N$ a normalization constant equal to
\begin{equation}
N = \sqrt{2 \sum_{l} (\lambda_{2l} + \lambda_{2l +1})^{2}}.
\end{equation}
With these values, we obtain
\begin{equation}
F_{MY}(\ket{\psi}) = |\braket{\psi}{\phi}|^{2} = \sum_{l} \frac{\left(\lambda_{2l} + \lambda_{2l + 1} \right)^{2}}{2}
\end{equation}

\section{Proof of $F_{MY} \leq F_{LO}$ with equality for pure states}\label{appb}

Let $\rho$ be given.  Then
\begin{equation}
F_{LO}(\rho) = \max_{\Phi \in LO} F(\Phi(\rho), \proj{\phi_{+}}{\phi_{+}})
\end{equation}
with $LO$ the set of local operations that take the space $AB$ to a pair of qubits.  We may restrict this set to operations which only apply local unitaries and trace out everything but a pair of qubits to obtain
\begin{equation}
F_{LO}(\rho) \geq \max_{U, V} F(\text{tr}_{X} (U \otimes V \rho U^{\dagger} \otimes V^{\dagger}), \proj{\phi_{+}}{\phi_{+}})
\end{equation}
where $\text{tr}_{X}$ means tracing out everything but a pair of qubits.  Since Fidelity only increases when a system is traced out we have
\begin{equation}
F_{LO}(\rho) \geq \max_{U,V} F(U \otimes V \rho U^{\dagger}Ê\otimes V^{\dagger}, \proj{\phi}{\phi} \otimes \proj{\phi_{+}}{\phi_{+}})
\end{equation}
for all $\ket{\phi}$, and in particular for the $\ket{\phi}$ which maximizes the expression and gives $F_{MY}(\rho)$.  Thus
\begin{equation}
F_{LO}(\rho) \geq F_{MY}(\rho)
\end{equation}

Now suppose that $\rho = \proj{\psi}{\psi}_{AB}$.  We may write an operation in LO as adding a pair of ancillas and a pair of target qubits, applying a pair of unitaries, and tracing out everything but the target qubits.  Thus
\begin{widetext}
\begin{equation}
F_{LO}(\ket{\psi}) = \max_{U, V} F(\text{tr}_{ABX_{a}X_{b}}(U \otimes V\ket{\psi}_{AB} \ket{00}_{X_{a} X_{b}} \ket{00}_{Y_{a}Y_{b}}), \proj{\phi_{+}}{\phi_{+}}_{Y_{a}Y_{b}}) 
\end{equation}

Applying Uhlmann's theorem, we obtain
\begin{equation}
F_{LO}(\ket{\psi}) = \max_{U, V, \ket{\phi}} \left|\bra{\psi}_{AB} \bra{00}_{X_{a} X_{b}} \bra{00}_{Y_{a}Y_{b}}U^{\dagger} \otimes V^{\dagger} \ket{\phi}\otimes\ket{\phi_{+}}_{Y_{a}Y_{b}}\right|^{2}  
\end{equation}
\end{widetext}
The right hand side is equal to $F_{MY}(\ket{\psi}\otimes \ket{00} \otimes \ket{00})$ by definition.  This in turn is equal to $F_{MY} (\ket{\psi})$ since the value of $F_{MY}$ for a pure state is only dependent on the Schmidt decomposition, which the product state ancillas do no change.  Thus
\begin{equation}
F_{LO}(\ket{\psi}) = F_{MY}(\ket{\psi})
\end{equation}

For mixed states there exist cases with a strict inequality.  For example $F_{MY}(\frac{I}{4}) = \frac{1}{4}$, but $F_{LO}(\frac{I}{4}) = \frac{1}{2}$ since the class $LO$ allows us to replace the state with $\ket{00}$.

\section{Proof of the lower bound for $F_{LO}$.}\label{appc}

Here we prove the bound on $F_{LO}$ eq. (\ref{fLObound}).
The definition of $F_{LO}$ is:
\begin{eqnarray}
F_{LO}&=&\max_{M_k , N_l} \sum_{k,l} \Tr \left[ M_k\otimes N_L \rho M_k^\dagger \otimes N_l^\dagger\   \boldsymbol{\Phi^+}\right]\nonumber\\
& &M_k: H_A \to \compl^2 \quad;\quad
\sum_k M_k^\dagger M_k =\one_A\nonumber\\
& &N_l: H_B \to \compl^2 \quad;\quad
\sum_l N_l^\dagger N_l =\one_B
\end{eqnarray}
where $\{M_k\}$ ($\{ N_l\}$) are CP maps from Alice (Bob's) system to 2 dimensional spaces. In general $\{M_k\}$ and $\{ N_l\}$ will depend on $\rho$.

We will explicitly describe CP map's that achieve eq. (\ref{fLObound}), thereby showing that it is a lower bound for $F_{LO}$. This bound is certainly not tight. This can be seen by the construction we use, since the CP maps depend on the measurement operators $\hat{A}$, $\hat{A}^\prime$, $\hat{B}$, $\hat{B}^\prime$, but not on the state $\rho$ itself. Thus the CP maps do not use all the available information, and cannot not be optimal. 

The CP maps are constructed as follows:
\begin{enumerate}
\item The operators $\hat{A}$, $\hat{A}^\prime$ (and $\hat{B}$, $\hat{B}^\prime$) are block diagonal, where each block is a $2\times 2$ matrix. We use the projectors $\Pi_{\alpha}\otimes\Pi_\beta$ to project onto these blocks, obtaining states $\rho_{\alpha\beta}$ with probability $p_{\alpha\beta}$. The Bell operator in block $(\alpha,\beta)$ has expectation $S(\rho_{\alpha\beta})$.
\item If $S(\rho_{\alpha\beta})\leq 2$, then $F\geq 0$. In this case do nothing.
\item If $S(\rho_{\alpha\beta}) >  2$, then 
carry out local rotations, such that after the rotations, the measurements look like
\begin{eqnarray}
\hat A&=&\cos a Z + \sin a X\nonumber\\
\hat A'&=&\sin a Z + \cos a X\nonumber\\
&|a|&\leq \frac{\pi}{4}
\label{AA'}
\end{eqnarray}
and
\begin{eqnarray}
\hat B&=&\cos(\frac{\pi}{4} + b) X + \sin (\frac{\pi}{4} + b) Z \nonumber\\
&=& \frac{1}{\sqrt{2}} (\cos b - \sin b) X + \frac{1}{\sqrt{2}} (\cos b + \sin b) Z\nonumber\\
\hat B'&=&-\cos (\frac{\pi}{4} + b) X + \sin (\frac{\pi}{4} + b) Z\nonumber\\
&=&-\frac{1}{\sqrt{2}} (\cos b - \sin b) X + \frac{1}{\sqrt{2}} (\cos b + \sin b) Z\nonumber\\
&|b|&\leq \frac{\pi}{4}
\label{BB'}
\end{eqnarray}
\end{enumerate}
The idea of the final rotations is that the operators $\hat{A}$, $\hat{A}^\prime$ (and $\hat{B}$, $\hat{B}^\prime$) define local bases, and we rotate the state so that these bases are aligned with the local bases defined by the state $\boldsymbol{\Phi^+}$.


Note that as both the fidelity and the CHSH violation are linear functions of the density matrix $\rho$, we can restrict ourselves to pure states.  Furthermore, because of the linearity of $S$ and $F$, we can focus on one block $(\alpha, \beta)$. Taking the concave hull will yield the set of accessible points. From now on we drop the indices $\alpha, \beta$.

Using eqs. (\ref{AA'}, \ref{BB'}) the Bell operator takes the form
\begin{eqnarray}
\hat B 
&=&\sqrt{2}\left[ \cos a \cos b (ZZ+XX) + \cos a \sin b (ZZ-XX)\right.\nonumber\\
& &\left.+\sin a \cos b (ZX+XZ)
+\sin a \sin b (-ZX+XZ)\right]
\nonumber
\end{eqnarray}
The eigenvectors of $\hat B$ are denoted $\ket{\Phi_i}$ and its eigenvectors are  $\lambda_1, \lambda_2, -\lambda_2, -\lambda_1$ where $0\leq \lambda_2\leq 2\leq \lambda_1\leq 2\sqrt{2}$.
Explicitly we have:
\begin{eqnarray}
\lambda_1(a,b)&=&2\sqrt{1+\cos 2a \cos 2b}\nonumber\\
&=&2\sqrt{2}\sqrt{\cos^2 a \cos^2 b + \sin^2a \sin^2 b}\nonumber\\
\lambda_2(a,b)&=&2\sqrt{1-\cos 2a \cos 2b}\end{eqnarray}
The fidelities of the eigenvectors with the $\ket{\Phi^+}$ state are $f_i=|\langle{\Phi^+} | \Phi_i\rangle|^2$.
One finds
\begin{eqnarray}
f_2=f_3&=&0\nonumber\\
f_1+f_4&=&1\nonumber\\
\Delta f = f_1-f_4&=&\frac{2 \sqrt{2} \cos a \cos b}{\lambda_1(a,b)}\label{Df}
\end{eqnarray}

The pure state on which the measurements are carried out can be written in the basis $\ket{\Phi_i}$ as:
$$\ket \psi = c_1 \ket {\Phi_1} + c_2 \ket {\Phi_2} + c_3 \ket {\Phi_3} + c_4 \ket {\Phi_4}\ . $$
The expectation of CHSH is 
$$\langle \hat B\rangle = S = (|c_1|^2-|c_4|^2) \lambda_1 + (|c_2|^2-|c_3|^2) \lambda_2 $$
and the Fidelity is
$$F=|c_1 \sqrt{f_1} + c_4 \sqrt{f_4}|^2$$ (where we take the $\sqrt{f_1}$ and $\sqrt{f_4}$ to be the postive square roots of $f_1$ and $f_4$).

Note that these expressions would be unchanged if we had a mixture of $\ket \Phi_2$, $\ket \Phi_3$, and $c_1 \ket \Phi_1  + c_4 \ket \Phi_4$. Henceforth we consider such a mixture.
If the state is of the form $\ket \psi = \ket \Phi_3$ or of the form
$\ket \psi = \ket \Phi_2$, then $S\leq 2$, and therefore, trivially, $F\geq 0$.

We now concentrate on the non trivial case 
$\ket \psi =c_1 \ket \Phi_1  + c_4 \ket \Phi_4$ and $2 < S \leq 2 \sqrt{2}$.
We will show that in this case
\begin{equation}
\frac{1}{2} + \frac{S}{4\sqrt{2}}\leq F\leq1\label{Phi1Phi2}
\end{equation}

Note that eq. (\ref{Phi1Phi2}) and the preceding arguments give us the extremal points in the $(S,F)$ plane we were searching for. Taking the concave hull yields eq. (\ref{fLObound}). The concave hull can be attained by taking the angles $a=b=0$, and as state a mixture of $\ket{\phi^+}$ (which has $F=1$ and $S=2\sqrt{2}$ and of $\ket{\Phi_2}$ (which has $F=0$ and $S=2$).

{\bf Proof of eq. (\ref{Phi1Phi2}).}

To prove eq. (\ref{Phi1Phi2}) this recall that
$F=|c_1 \sqrt{f_1} + c_4 \sqrt{f_4}|^2$ and $S=(|c_1|^2-|c_4|^2) \lambda_1$.
We can view $F=|\vec v\cdot \vec w|^2$ as the scalar product of two vectors $\vec v = (c_1, c_4)$ and $\vec w= (\sqrt{f_1}, \sqrt{f_4})$. 
Our argument will be to fix $S$ and to minimize $F$.

For fixed $S$, $a$, $b$, the minimum of $F$ is obtained when
\begin{eqnarray}
c_1= + \sqrt{\frac{1}{2} + \frac{S}{2 \lambda_1}}\quad , \quad
c_4= - \sqrt{\frac{1}{2} - \frac{S}{2 \lambda_1}}.
\end{eqnarray}
From now on we take $c_1, c_4$ to have this form.

We can then write
\begin{equation}
F=\left(
 \sqrt{\frac{1}{2} + \frac{S}{2 \lambda_1}}
  \sqrt{\frac{1}{2} + \frac{\Delta f}{2}}
-   \sqrt{\frac{1}{2} - \frac{S}{2 \lambda_1}}
    \sqrt{\frac{1}{2} - \frac{\Delta f}{2}}
    \right)^2
    \label{FFF}
\end{equation}
From now on, our aim is to choose the measurement angles $a,b$ that minimize eq. (\ref{FFF}) for fixed $S$.

First let us keep $S$ and $\lambda_1$ fixed. Then $F$ is minimum when $\Delta f$ is minimized. We show that this occurs when $|a|=|b|$.

{\bf Proof.}
We consider the $(a,b)$ plane. The vector
$$\vec n =(  -\cos a \sin a (\cos^2b - \sin^2b), -\cos b \sin b (\cos^2a - \sin^2a))$$
is normal to the surfaces $\lambda_1=constant$; and the vector
$$\vec t =( \cos b \sin b (\cos^2a - \sin^2a), -\cos a \sin a (\cos^2b - \sin^2b)$$ is tangent to the surfaces $\lambda_1= constant$.

Recall equation (\ref{Df}). It then follows that
$\vec t \cdot (-\sin a \cos b, -\cos a \sin b)=\sin a \sin b (\sin^2a \cos^2b - \cos^2a \sin^2b)$ is proportional to the change of $\Delta f$ along the surfaces $\lambda_1=constant$. Analyzing this function, one finds that the minimum of $\Delta f$ occurs when $|a|=|b|$.
{\bf End of proof.}

We can thus replace $|a|=|b|$ in eq. (\ref{FFF}). 
Then when $S>2$ the minimum of $F$ occurs when $a=b=0$. Replacing $a=b=0$ in eq. (\ref{FFF}), this is equivalent to proving that when $|a|=|b|$, $F\geq \frac{1}{2} + \frac{S}{4\sqrt{2}}$.

{\bf Proof.} 
We use eq. (\ref{FFF}) to rewrite the inequality $F\geq \frac{1}{2} + \frac{S}{4\sqrt{2}}$ as
\begin{equation}
\frac{1}{2} + \frac{S \Delta f }{2 \lambda_1}
-2 \sqrt{\frac{1}{4} - \frac{S^2}{4 \lambda_1^2}}
  \sqrt{\frac{1}{4} - \frac{\Delta f^2}{4}}
\geq 
\frac{1}{2} + \frac{S}{4\sqrt{2}}
\end{equation}
which we reorganise as
\begin{equation}
 \frac{S \Delta f }{2 \lambda_1}-\frac{S}{4\sqrt{2}} \geq
2 \sqrt{\frac{1}{4} - \frac{S^2}{4 \lambda_1^2}}
  \sqrt{\frac{1}{4} - \frac{\Delta f^2}{4}} \ .
\end{equation}
Both the left hand side and the right hand side are positive (since it is easily checked that
$\Delta f /S_1 \geq 1/2\sqrt{2}$). Hence this inequality is equivalent to its square, which gives:
\begin{equation}
-\frac{S^2 \Delta f }{\sqrt{2} \lambda_1}
+\frac{S^2}{8}\geq 1 - \Delta f^2 - \frac{S^2}{\lambda_1^2} \ .
\end{equation}
Reorganising terms yields
\begin{equation}
\frac{2  (\cos a^2 -1)^2 (S^2-4)}{\lambda_1^2}\geq 0
\end{equation}
which is manifestly true when $S\geq 2$.
{\bf End of proof.}

{\bf End of proof of eq. (\ref{Phi1Phi2}).}

\end{document}